\documentclass[fleqn,usenatbib]{mnras}

\usepackage{newtxtext,newtxmath}

\usepackage[T1]{fontenc}

\DeclareRobustCommand{\VAN}[3]{#2}
\let\VANthebibliography\thebibliography
\def\thebibliography{\DeclareRobustCommand{\VAN}[3]{##3}\VANthebibliography}

\usepackage{graphicx}	
\usepackage{amsmath}	

\title[Polar alignment of massive retrograde circumbinary discs]{Polar alignment of a massive retrograde circumbinary disc around an eccentric binary}

\author[C. P. Abod et al.]{Charles P. Abod,$^{1}$\thanks{E-mail: abod@unlv.nevada.edu}
Cheng Chen,$^{1}$
Jeremy Smallwood,$^{1}$
Ian Rabago,$^{1}$
Rebecca G. Martin,$^{1}$\newauthor
and Stephen H. Lubow$^2$
\\$^1$Department of Physics and Astronomy,  University of Nevada, Las Vegas, 4505 South Maryland Parkway, Las Vegas, NV 89154, USA 
\\$^{2}$Space Telescope Science Institute, 3700 San Martin Drive, Baltimore, MD 21218, USA\\}

\date{Accepted XXX. Received YYY; in original form ZZZ}

\pubyear{2021}

\begin{document}
\label{firstpage}
\pagerange{\pageref{firstpage}--\pageref{lastpage}}
\maketitle

\begin{abstract}
A test particle orbit around an eccentric binary has two stationary states in which there is no nodal precession: coplanar and polar. Nodal precession of a misaligned test particle orbit centres on one of these stationary states. A low mass circumbinary disc undergoes the same precession and moves towards one of these states through dissipation within the disc.  For a massive particle orbit, the stationary polar alignment occurs at an inclination less than $90^{\circ}$, this is the prograde-polar stationary inclination. A sufficiently high angular momentum particle has an additional higher inclination stationary state, the retrograde-polar stationary inclination. Misaligned particle orbits close to the retrograde-polar stationary inclination are not nested like the orbits close to the other stationary points. We investigate the evolution of a gas disc that begins close to the retrograde-polar stationary inclination. With hydrodynamical disc simulations, we find that the disc moves through the unnested crescent shape precession orbits and eventually moves towards the prograde-polar stationary inclination thus increasing the parameter space over which circumbinary discs move towards polar alignment. If protoplanetary discs form with an isotropic orientation relative to the binary orbit, then polar discs may be more common than coplanar discs around eccentric binaries, even for massive discs. This has implications for the alignment of circumbinary planets.
\end{abstract}



\section{Introduction}

Binary star systems are commonly observed \citep[e.g.][]{Ghez_1993,Duchene2013}. They are known to form in turbulent molecular clouds \citep[]{McKee2007} where the accretion process may be chaotic \citep[][]{Bate2003,Bate2018}. This results in a high likelihood that the protoplanetary disc that forms is misaligned with respect to the orbital plane of the binary \citep[][]{Monin2007,Bateetal2010,Bate2018}. Circumbinary disc observations suggest that misalignments are common \citep[e.g.][]{Winn2004,Chiang2004,Capelo2012,Brinch2016,Kennedy2012,Aly2018}. Circumbinary discs have been observed with polar inclinations of $90^\circ$ to the binary orbit. HD98800 and V773 Tau B have polar circumbinary gas discs \citep{Kennedy2019,Kenworthy2022} while 99 Her is a polar circumbinary debris disc \citep{Kennedy2012,Smallwood2020}. The formation and evolution of planets around binaries may be altered by the torque from the binary \citep[e.g.][]{Nelson2000,Mayer2005,Boss2006,Martinetal2014,Fu2015,Fu2015b,Fu2017,Franchini2019}. Giant planets that form in a misaligned disc may not remain coplanar to the disc \citep[][]{Picogna2015,Lubow2016,Martin2016}. We seek to understand the evolution of misaligned circumbinary discs in order to better understand the observed properties of exoplanets.

The Kepler Mission has so far detected about 20 circumbinary planets \citep[e.g.][]{Doyle2011,Welshetal2012,Orosz2012,Kostov2016} and the TESS Mission recently discovered one more \citep[][]{Kostov2021}. All of the circumbinary planets detected so far are nearly coplanar to the binary orbital plane. However, this may be a selection effect due to the small orbital period of the Kepler binaries \citep[][]{Czekala2019}. Longer orbital period binaries are  expected to host planets with a wide range of inclinations. Terrestrial circumbinary planets that form through core accretion in the absence of gas may only form in polar or coplanar configurations \citep{Childs2021,Childs2021b,Childs2022}, but gas giants that form within the gas disc may form with the same initial inclination as the gas disc \citep[e.g.][]{Pierens2018}.   Planets with large misalignments are much more difficult to detect than those that orbit in the binary orbital plane because their transits are much rarer \citep{Schneider1994,MartinD2014,MartinD2015,MartinD2017,Chen2021}. However, other detection methods are possible, such as eclipse timing variations \citep{Murat2022}. Polar planets (those in an orbit close to perpendicular to the binary orbital plane) may be distinguished from coplanar planets through eclipse timing variations of the binary \citep[][]{Zhang2019}. Two circumbinary planets have been found around eccentric orbit binaries. Kepler-34b has a mass of $0.22 M_{\rm J}$ and orbits an eclipsing binary star system (Kepler-34) that has an orbital eccentricity of 0.52 \citep[][]{Welshetal2012,Kley2015}.

The evolution of low mass circumbinary discs is fairly well understood. A misaligned circumbinary disc around a circular orbit binary experiences uniform nodal precession with constant tilt in the absence of dissipation  \citep[e.g.][]{Nixon2012,Facchinietal2013,Lodato2013,Foucart2013}. The angular momentum vector of the disc precesses about the angular momentum vector of the binary. We call this a {\it circulating} solution. In this work we focus on protoplanetary discs that are in the wave–like regime in which the disc aspect ratio is much larger than the \cite{SS1973} viscosity parameter, $H/R > \alpha$. For a sufficiently warm and radially narrow disc, the disc precesses as a solid body \citep[e.g.][]{Larwoodetal1997}. In these systems, dissipation in the disc leads to alignment with the binary orbital plane \citep[][]{Papaloizouetal1995,Lubow2000,Nixonetal2011b,Nixon2012,Facchinietal2013,Lodato2013,Foucart2013,Foucart2014}. 

The evolution around an eccentric orbit binary is more complex. For a sufficiently high initial tilt, a low mass disc precesses about the eccentricity vector of the binary (rather than the orbital rotation axis of the binary) \citep{Verrier2009,Farago2010,Doolin2011,Aly2015}. We call this a {\it librating} solution, where the orbit is precessing about the {\it prograde-polar} stationary state. In the low mass disc case, the prograde-polar stationary state is aligned with the binary eccentricity vector. Dissipation leads to polar alignment where the circumbinary disc is aligned to the eccentricity vector of the binary and perpendicular to the binary orbital plane \citep{Martin2017,Lubow2018,Zanazzi2018,Martin2018,Cuello2019}. 

The evolution of a {\it massive} circumbinary gas disc around an eccentric binary has not yet been fully explored. In the absence of dissipation within the disc, the dynamical evolution of a radially narrow circumbinary gas disc is qualitatively similar that of a massive particle orbit around the binary. Thus in order to understand the evolution of a massive disc, in Section~\ref{section:threebody}, we first explore the dynamics of massive particle orbits \citep[see also][]{Chen_2019}. In this case, the binary feels the gravitational force of the planet which causes the binary orbit to evolve. The eccentricity vector of the binary precesses, the binary orbit tilts and the magnitude of its eccentricity oscillates. The prograde-polar stationary aligned state occurs at a lower level of misalignment \citep{Farago2010,Zanazzi2018,Martin_2019, Chen_2019}. For librating orbits, the angular momentum vector precesses around the prograde-polar stationary state. For a gas disc, dissipation causes these oscillations to damp and the disc settles at the prograde-polar stationary tilt angle. 

In the massive particle case, there is a third type of nodal precession orbit, that we call the {\it crescent} orbits (\citet{Martin_2019,Chen2019} see later section~\ref{sec:mid_threebody} and Fig.~\ref{fig:big_array} for more details.). These occur for sufficiently massive particles that begin closer to retrograde than to prograde. These crescent orbits are not nested like the circulating and librating orbits, meaning that they do not have a common centre for the precession. For a sufficiently high angular momentum particle, there is an additional {\it retrograde-polar} stationary inclination in the region of these crescent orbits. The particle precession in a crescent orbit is not centred around a stationary state. As we show in this paper, a gas disc on such orbits often evolves towards the prograde-polar stationary tilt angle. In Section~\ref{sec:circumbinary} we consider the long term evolution of a gas disc that begins in this regime, close to the retrograde-polar stationary inclination.  Finally, in Section \ref{sec:conclusion}, we discuss the implications of our results and state our conclusions.

\section{Three-Body Simulations}
\label{section:threebody}

\begin{table*}
	\centering
	\caption{This table lists the simulation parameters and outcomes. Each line consists of both an $n$-body planet simulation and an SPH disc simulation with the same combination of $j$ and $i_0$. The first column describes the simulation name. The second column is the initial inclination of the planet/disc. The third column is the angular momentum ratio of the planet/disc to the binary. The fourth column is the analytically calculated retrograde-polar stationary inclination (see Equation~(\ref{eq:crit_angle})). The fifth column is the mass of the planet. The sixth column describes whether the orbit of the planet is retrograde circulating (C$_{\rm R}$), librating (L) or in a crescent orbit (CRESCENT). The seventh column is the mass of the disc. The eighth column describes the orbit type of the disc prior to its $30 \%$ mass loss cut off. The ninth column describes the disc break radius. The tenth column is the time at which the circumbinary disc simulation has lost $30 \%$ of its initial mass.}
	\begin{tabular}{lccccccccccr}
		Name & $i_0/^{\circ}$ & $j$ & $i_{\rm rs}/^{\circ}$ & $M_{\rm p}/M$ & Planet Orbit & $M_{\rm d}/M$ & Disc Orbit & Break Radius/$a_{\rm b}$ & Time ($T_{\rm b}$)\\
		\hline
		\hline
		Low-$j$-120 & 120 & 0.5 & - & 0.006 & L & 0.01 & L & - & 2110\\
		Low-$j$-130 & 130 & 0.5 & - & 0.006 & L & 0.01 & L & - & 1540\\
		Low-$j$-140 & 140 & 0.5 & - & 0.006 & L & 0.01 & L & - & 1070\\
		Low-$j$-150 & 150 & 0.5 & - & 0.006 & L & 0.01 & L & - & 850\\
		Low-$j$-160 & 160 & 0.5 & - & 0.006 & L & 0.01 & L & 6 & 1560\\
		Low-$j$-170 & 170 & 0.5 & - & 0.006 & L & 0.01 & L & - & 2520\\
		\hline
		Mid-$j$-100 & 100 & 1.5 & 139 & 0.018 & L & 0.03 & L & - & 2880\\
		Mid-$j$-110 & 110 & 1.5 & 139 & 0.018 & CRESCENT & 0.03 & L & - & 2320\\
		Mid-$j$-120 & 120 & 1.5 & 139 & 0.018 & CRESCENT & 0.03 & CRESCENT $\rightarrow$ L & - & 1830\\
		Mid-$j$-130 & 130 & 1.5 & 139 & 0.018 & CRESCENT & 0.03 & CRESCENT & - & 1410\\
		Mid-$j$-140 & 140 & 1.5 & 139 & 0.018 & $i_{\rm rs}$ & 0.03 & $i_{\rm rs}$ $\rightarrow$ L & - & 1140\\
		Mid-$j$-150 & 150 & 1.5 & 139 & 0.018 & CRESCENT & 0.03 & CRESCENT (?) & - & 870\\
		Mid-$j$-160 & 160 & 1.5 & 139 & 0.018 & C$_{\rm R}$ & 0.03 & C$_{\rm R}$ & 2.5 & 1000\\
		Mid-$j$-170 & 170 & 1.5 & 139 & 0.018 & C$_{\rm R}$ & 0.03 & C$_{\rm R}$ & - & 2620\\
		\hline
		High-$j$-100 & 100 & 2.5 & 129 & 0.031 & L & 0.05 & L & - & 2380\\
		High-$j$-110 & 110 & 2.5 & 129 & 0.031 & CRESCENT & 0.05 & CRESCENT $\rightarrow$ L & - & 2310\\
		High-$j$-120 & 120 & 2.5 & 129 & 0.031 & CRESCENT & 0.05 & CRESCENT & - & 2230\\
		High-$j$-130 & 130 & 2.5 & 129 & 0.031 & $i_{\rm rs}$ & 0.05 & $i_{\rm rs}$ $\rightarrow$ L & - & 4200\\
		High-$j$-140 & 140 & 2.5 & 129 & 0.031 & CRESCENT & 0.05 & CRESCENT $\rightarrow$ L & - & 1980\\
		High-$j$-150 & 150 & 2.5 & 129 & 0.031 & C$_{\rm R}$ & 0.05 & C$_{\rm R}$ & - & 2020\\
		High-$j$-160 & 160 & 2.5 & 129 & 0.031 & C$_{\rm R}$ & 0.05 & C$_{\rm R}$ & 2.5 & 1110\\
		High-$j$-170 & 170 & 2.5 & 129 & 0.031 & C$_{\rm R}$ & 0.05 & C$_{\rm R}$ & - & 2620\\
		\hline
	\end{tabular}
		\label{tab:sim_list}
\end{table*}

In this Section, before running our circumbinary disc simulations, we first consider the evolution of a three-body system in order to understand the dynamics of massive circumbinary particle orbits. The parameters we vary are the initial inclination of the third body/planet orbit with respect to the inner binary orbital plane, $i_0$, and the angular momentum ratio of the third body compared to the inner binary, $j$. Note that in this paper, $j$ refers to the angular momentum ratio between both planet-binary and disc-binary for the three-body systems and circumbinary disc systems, respectively. For the three-body simulations, we follow the methods outlined in \cite{Chen_2019}. In Section~\ref{sec:circumbinary} we compare these results with circumbinary disc simulations which have roughly the same initial parameters, $j$ and $i_0$.

\subsection{Three-body simulation set-up}
\label{sec:threebody_methods}

We use the $n$-body simulation package, {\sc REBOUND}, with the WHfast integrator which is a second order symplectic Wisdom Holman integrator with 11th order symplectic correctors \citep[][]{rein2015}. We solve the gravitational equations for a planet orbiting around a binary star system. 
The primary mass in the binary is set to be $M_{1} = 0.9\,M$, while the secondary mass is $M_{2} = 0.1\,M$, where $M=M_1+M_2$ is the total mass of the binary. The initial eccentricity of the binary orbit is set to $e_{\rm b} = 0.8$, and the initial semi-major axis of the binary is $a_{\rm b}$.

The third body is a planet with mass $M_{\rm p}$ that has an initially circular, Keplerian orbit around the centre of mass of the binary. The six orbital elements that define the trajectory of the planet are: its semi-major axis $a_{\rm p}$, inclination $i$, eccentricity $e_{\rm p}$, longitude of the ascending node $\phi$, argument of periapsis $\omega$, and true anomaly $\nu$. Initially we set $e_{\rm p} = 0$ and $\omega = 0$ since the planet's orbit is initially circular. We also set $\nu = 0$ and $\phi = 90^{\circ}$. The binary orbit is not fixed and can feel the gravity of the orbiting planet. Therefore, to remove the chance of running into orbital instability, we follow \cite{Chen_2019} and choose a relatively large semi-major axis of the planet of $a_{\rm p} = 20 \,a_{\rm b}$, where the orbits are stable for all inclinations \citep{Chen2020}. Consequently, this increases the time required to complete the three-body simulations, which we take to be $6000\,T_{\rm b}$, where the binary orbital period is $T_{\rm b}$.  We define $j$ to be the planet-to-binary angular momentum ratio. For our three-body simulations, 
\begin{equation}
j = \frac{J_{\rm p}}{J_{\rm b}},
\end{equation}
where $J_{\rm p}$ is the angular momentum of the orbiting planet given by
\begin{equation}
    J_{\rm p}=M_{\rm p} \sqrt{ G(M_1+M_2+M_{\rm p}) a_{\rm p}}
\end{equation}
and $J_{\rm b}$ is the angular momentum of the binary given by
\begin{equation}
    J_{\rm b}=\frac{M_1M_2}{M_1+M_2} \sqrt{G (M_1+M_2) a_{\rm b}(1-e_{\rm b}^2)}.
\end{equation}
We vary the planet mass to give the required value for $j$. Note that the system dynamics depend only on the angular momentum ratio $j$ and the binary eccentricity $e_{\rm b}$ \citep[][]{Martin_2019}. We probe systems with three different $j$ values, each with a set of different initial inclinations, $i_0$, see Table~\ref{tab:sim_list}. The planet's orbit remains approximately circular throughout the simulations, as is expected analytically since the particle eccentricity is a constant of motion in the secular quadrupole approximation for the binary \citep[][]{Farago2010}.

The stationary inclinations for the test particle occur when the particle orbit displays no precession with respect to the binary. These can be calculated analytically with
\begin{equation}
    \cos{i}=\frac{-(1+4e_{\rm b}^{2}) \pm \sqrt{(1+4e_{\rm b}^{2})^{2} + 60(1-e_{\rm b}^{2})j^{2}}}{10j}
	\label{eq:incstat} 
\end{equation}
\citep{Martin_2019,Chen_2019}. (Note that there is a typo in Equation~5 in \cite{Chen_2019}.) We take the positive sign to calculate the prograde-polar stationary inclination, $i_{\rm s}$. This is the centre of the librating region. We take the negative sign in Equation~(\ref{eq:incstat}) to calculate the retrograde-polar stationary inclination $i_{\rm rs}$. But, this only exists ($\cos i\ge -1$) if the angular momentum ratio is greater than the critical value of
\begin{equation}
    j_{\rm cr}=\frac{1+4e_{\rm b}^{2}}{2+3e_{\rm b}^{2}}
	\label{eq:crit_angle}
\end{equation}
\citep[see Appendix A of][]{Martin_2019}. For the binary eccentricity of $e_{\rm b}=0.8$ considered in this work, $j_{\rm cr}=0.91$.

Given this value of $j_{\rm cr}$, we set our "low", "middle", and "high" values to be  $j = 0.5$, $1.5$, and $2.5$ respectively. We chose these values to probe above and below $j_{\rm cr}$. We choose two values greater than $j_{\rm cr}=0.91$ to further investigate the parameter regime where crescent orbits are found. We increase the initial angular momentum ratio between the planet and binary by increasing the planet mass.
The angular momentum ratio is given by
\begin{equation}
    j=\frac{\frac{M_{\rm p}}{M} \sqrt{\frac{a_{\rm p}}{a_{\rm b}}} \sqrt{1+\frac{M_{\rm p}}{M}}}{f_{\rm b}(1-f_{\rm b}) \sqrt{1-e_{\rm b}^2}},
	\label{eq:rebound_mass_j}
\end{equation}
where $f_{\rm b}=M_2 / M$ is the binary mass fraction.  We solve this for $M_{\rm p}$ values that are associated with $j = 0.5$, $1.5$, and $2.5$ and the values are shown in Table~\ref{tab:sim_list}.

\begin{figure*}
	\includegraphics[width=2\columnwidth]{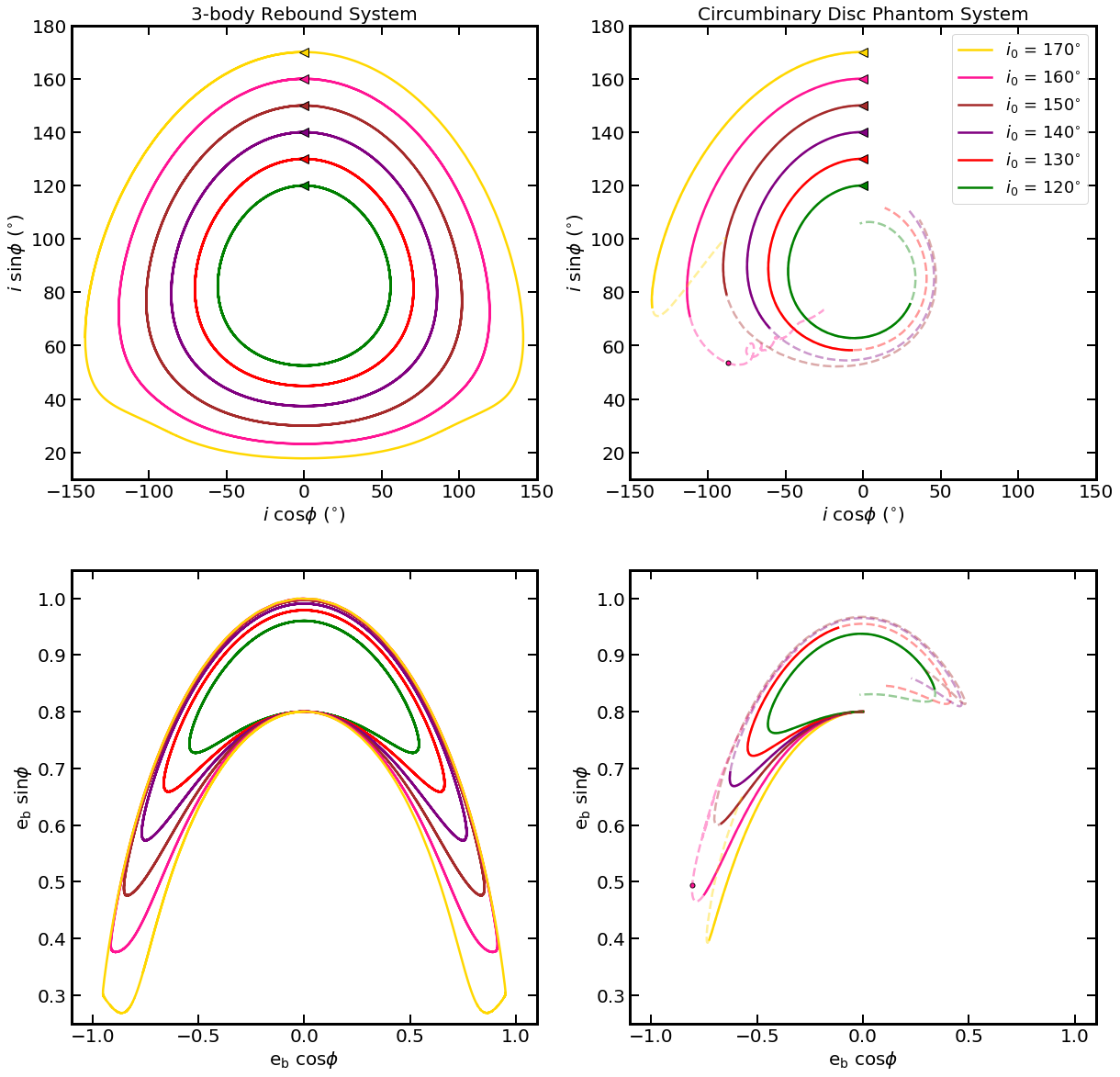}
    \caption{The upper panels show the $i\cos{\phi} - i\sin{\phi}$ phase plots while the lower panels show the $e_{\rm b}\cos{\phi} - e_{\rm b}\sin{\phi}$ phase plane. The left panels are for the three-body systems (see Section~\ref{section:threebody}) while the right panels are for  the circumbinary disc systems (see Section~\ref{sec:circumbinary}). The dashed lines represent any part of the system that is simulated past the point where the disc has lost $30 \%$ of its initial mass (see Fig.~\ref{fig:mass_loss}). All simulations, both disc and planet, have $j=0.5$ initially. The three-body system has a planet with mass $M_{\rm p} = 0.006 \,M$ and initial planet distance from system centre of mass of $r=20 \,a_{\rm b}$. The initial mass of the disc is $M_{\rm d} = 0.01 \,M$. Additionally, the pink dot on each of the right plots indicates the point where that circumbinary disc broke apart at time $T=2600 \,T_{\rm b}$.}
    \label{fig:md01}
\end{figure*}

\begin{figure*}
	\includegraphics[width=2\columnwidth]{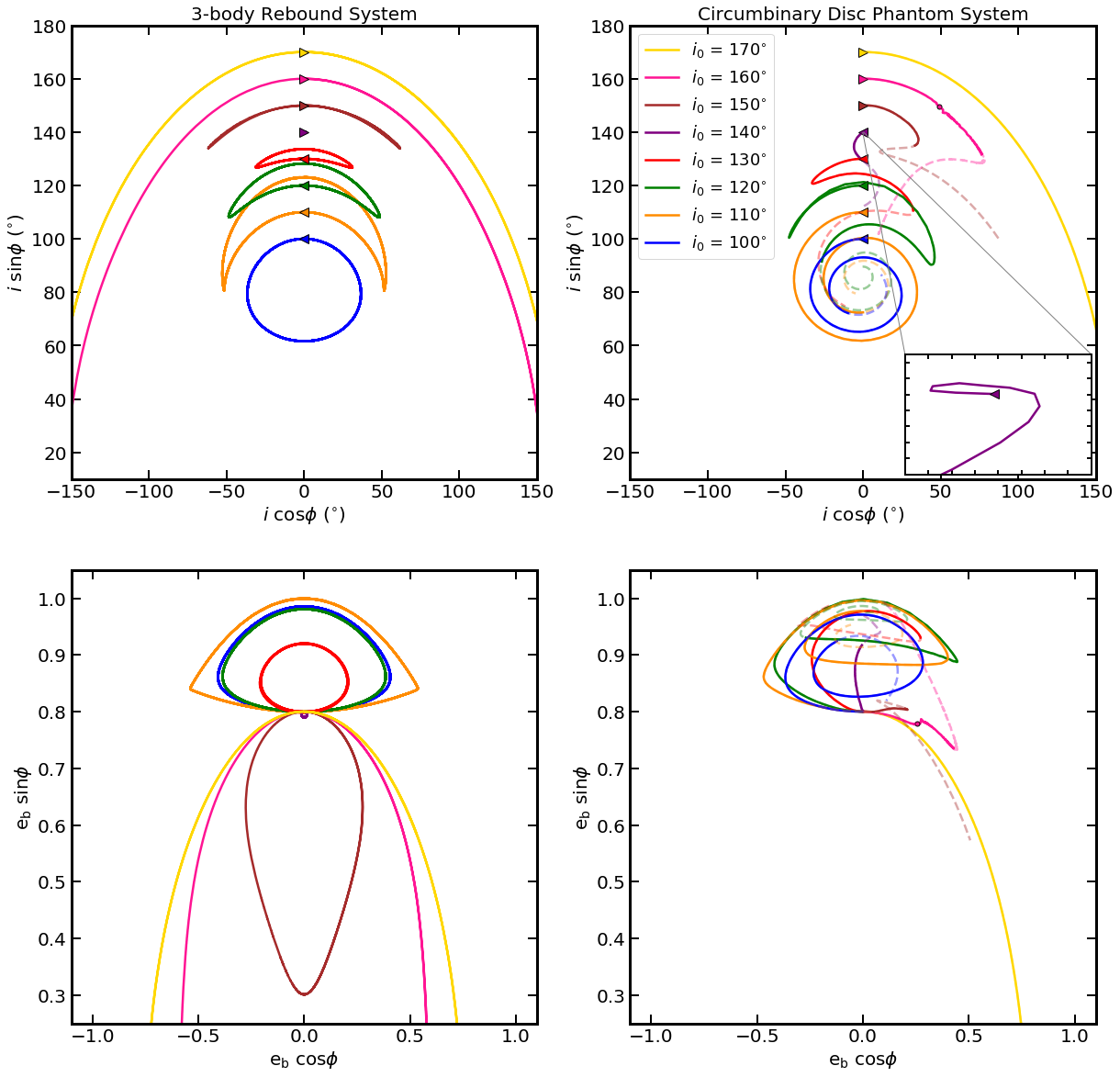}
    \caption{Same as Fig.~\ref{fig:md01} except $j=1.5$. The three-body systems have a planet with mass $M_{\rm p} = 0.018 \,M$ and initial planet distance from system centre of mass $r=20 \,a_{\rm b}$. The initial mass of the disc is $M_{\rm d} = 0.03 \,M$. Additionally, the pink dot on each of the right plots indicates the point where that circumbinary disc broke apart at $T=240 \,T_{\rm b}$.}
    \label{fig:md03}
\end{figure*}

\begin{figure*}
	\includegraphics[width=1.5\columnwidth]{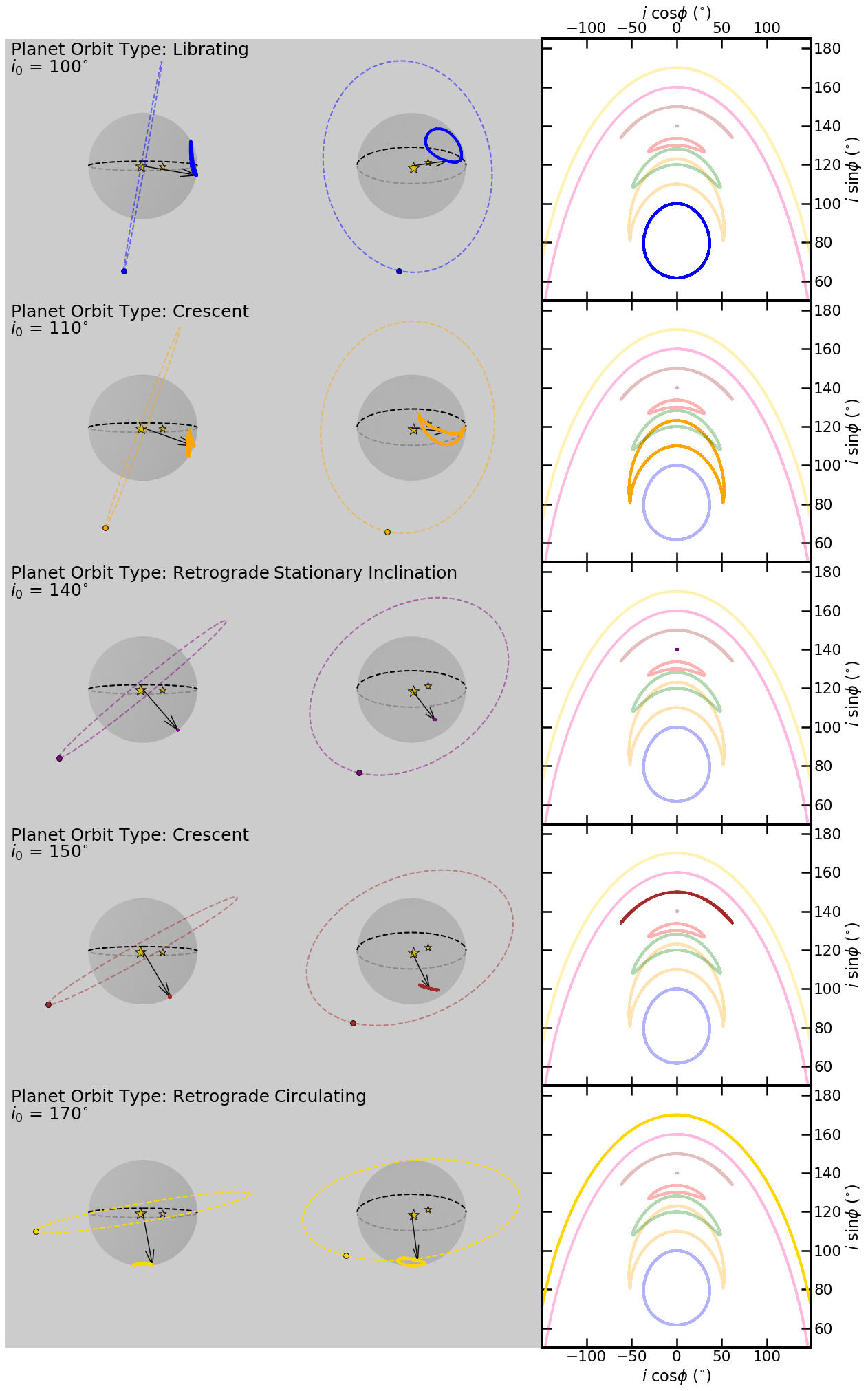}
    \caption{Precession paths in 3D for the Mid-$j$ simulations. Each row corresponds to a different initial inclination, $i_0$, that is highlighted in the inclination phase plot on the right that is the same as the upper left panel in Fig.~\ref{fig:md03}. The left and middle columns show 3D visual representations of the three-body simulations. The left column is viewed from azimuth$\,=0^{\circ}$ and elevation$\,=-5^{\circ}$, while the middle column is viewed from azimuth$\,=50^{\circ}$ and elevation$\,=-20^{\circ}$. The two yellow stars in the centre are the binary stars that are shown at apastron, the dotted coloured line and circle outside of the gray sphere are the third body with its trajectory, the black arrow is the angular momentum vector of the third body, and the solid coloured line on the sphere is the path of the planet's angular momentum vector for a complete precession. The black and grey dotted lines on the sphere show the plane in which the binary orbits.}
    \label{fig:big_array}
\end{figure*}

We work in a frame defined by the instantaneous values of the binary eccentricity vector, $\boldsymbol{e}_{\rm b}$, and angular momentum vector, $\boldsymbol{l}_{\rm b}$. The binary frame has three axes $\boldsymbol{e}_{\rm b}$, $\boldsymbol{l}_{\rm b}$, and $\boldsymbol{l}_{\rm b} \times \boldsymbol{e}_{\rm b}$. The inclination of the planet's orbital plane relative to the binary orbital plane is
\begin{equation}
    i=\cos^{-1}(\hat{\boldsymbol{l}}_{\rm b} \cdot \hat{\boldsymbol{l}}_{\rm p}),
	\label{eq:inc_bp}
\end{equation}
where the planet's angular momentum vector is $\boldsymbol{l}_{\rm p}$ and $\hat{}$ denotes a unit vector. The nodal phase angle of the planet is
\begin{equation}
    \phi=\frac{\pi}{2}+\tan^{-1}{\left(\frac{\hat{\boldsymbol{l}}_{\rm p} \cdot (\hat{\boldsymbol{l}}_{\rm b} \times \hat{\boldsymbol{e}}_{\rm b})}{\hat{\boldsymbol{l}}_{\rm p} \cdot \hat{\boldsymbol{e}}_{\rm b}}\right)}
	\label{eq:phi}
\end{equation}
\citep{Chen_2019,Chen2020err}.
In the following subsections, we explore the planet orbits produced from the three initial angular momentum ratio systems, each with their set of initial inclinations.

\subsection{Low-$j$ three-body system}
\label{sec:low_threebody}

We first consider three-body simulations with a relatively low third body angular momentum of  $j=0.5<j_{\rm cr}=0.91$. The prograde-polar stationary inclination calculated with Equation~(\ref{eq:incstat}) is $i_{\rm s}=82^{\circ}$. There is no retrograde-polar stationary inclination in this case since the angular momentum ratio is lower than the critical defined by Equation~(\ref{eq:crit_angle}). The two left panels in Fig.~\ref{fig:md01} show the results of the Low-$j$ three-body simulations. The top left plot shows the inclination phase plot, $i\cos{\phi} - i\sin{\phi}$. In this low mass case, all of the orbits shown in the phase plot are librating, meaning the particle angular momentum vector is precessing about the generalised prograde-polar inclination. All of the orbits begin at their respective initial inclinations and carry out their paths in a counterclockwise direction. The coloured triangle is located at the start of the path and points to the initial direction. The lower left panel shows the eccentricity phase plot $e_{\rm b}\cos \phi-e_{\rm b}\sin \phi$. All of the results in the eccentricity phase plot begin at $e_{\rm b} = 0.8$, and $\sin{\phi}=1$ and carry out their paths in a clockwise direction.
In all cases, the binary eccentricity oscillates and initially increases from its initial value.

\subsection{Mid-$j$ three-body system}
\label{sec:mid_threebody}

We now consider a more massive planet such that the angular momentum ratio $j=1.5>j_{\rm cr}=0.91$. The prograde-polar stationary inclination calculated with Equation~(\ref{eq:incstat}) is $i_{\rm s}=73^{\circ}$ and the retrograde-polar stationary inclination is $i_{\rm rs}=139^{\circ}$. The left two panels of Fig.~\ref{fig:md03} show the results of the Mid-$j$ three-body simulations. The top phase plot shows that for initial inclination $i_0 \lesssim 100^{\circ}$, the orbit is librating about the prograde-polar stationary inclination. For $i_0 \gtrsim 160^{\circ}$, the orbits are in retrograde-polar circulation, meaning that the planet angular momentum vector is precessing about the negative of the binary angular momentum vector. In the approximate initial inclination range, $110^{\circ}-150^{\circ}$ we see crescent orbits. These crescent orbits are not nested on a common centre like the circulating or librating orbits. The Mid-$j$-140 three-body simulation does not show significant variation from its initial inclination in the phase plot because of its proximity to the retrograde-polar stationary inclination at $i_{\rm rs}=139^{\circ}$. Additionally, as seen by the coloured triangles, the three-body systems with $i_0<i_{\rm rs}$ precess counterclockwise (prograde), while those with $i_0>i_{\rm rs}$ precess clockwise (retrograde).

Looking at the lower left panel in Fig.~\ref{fig:md03} in  we see that the binary eccentricity oscillates in all cases, but the initial direction varies depending upon the initial inclination. If $i_0<i_{\rm rs}$, then the eccentricity initially increases, while for $i_0> i_{\rm rs}$, the eccentricity initially decreases. The initial horizontal direction of the  paths in the inclination phase plot correlates with the initial vertical direction of the eccentricity plots. The orbital paths that have rightward-facing triangles have an initially decreasing $e_{\rm b}$, while orbits with leftward-facing triangles have an initially increasing $e_{\rm b}$.

To better understand what the phase plots represent in three-dimensions, Fig.~\ref{fig:big_array} connects the angular momentum evolution to the phase plots in Fig.~\ref{fig:md03}. From this depiction, we are able to see the true shapes of these orbital paths without the alterations that are made when mapping the spherical path onto a Cartesian coordinate system.

\subsection{High-$j$ three-body system}
\label{sec:high_threebody}

\begin{figure*}
	\includegraphics[width=2\columnwidth]{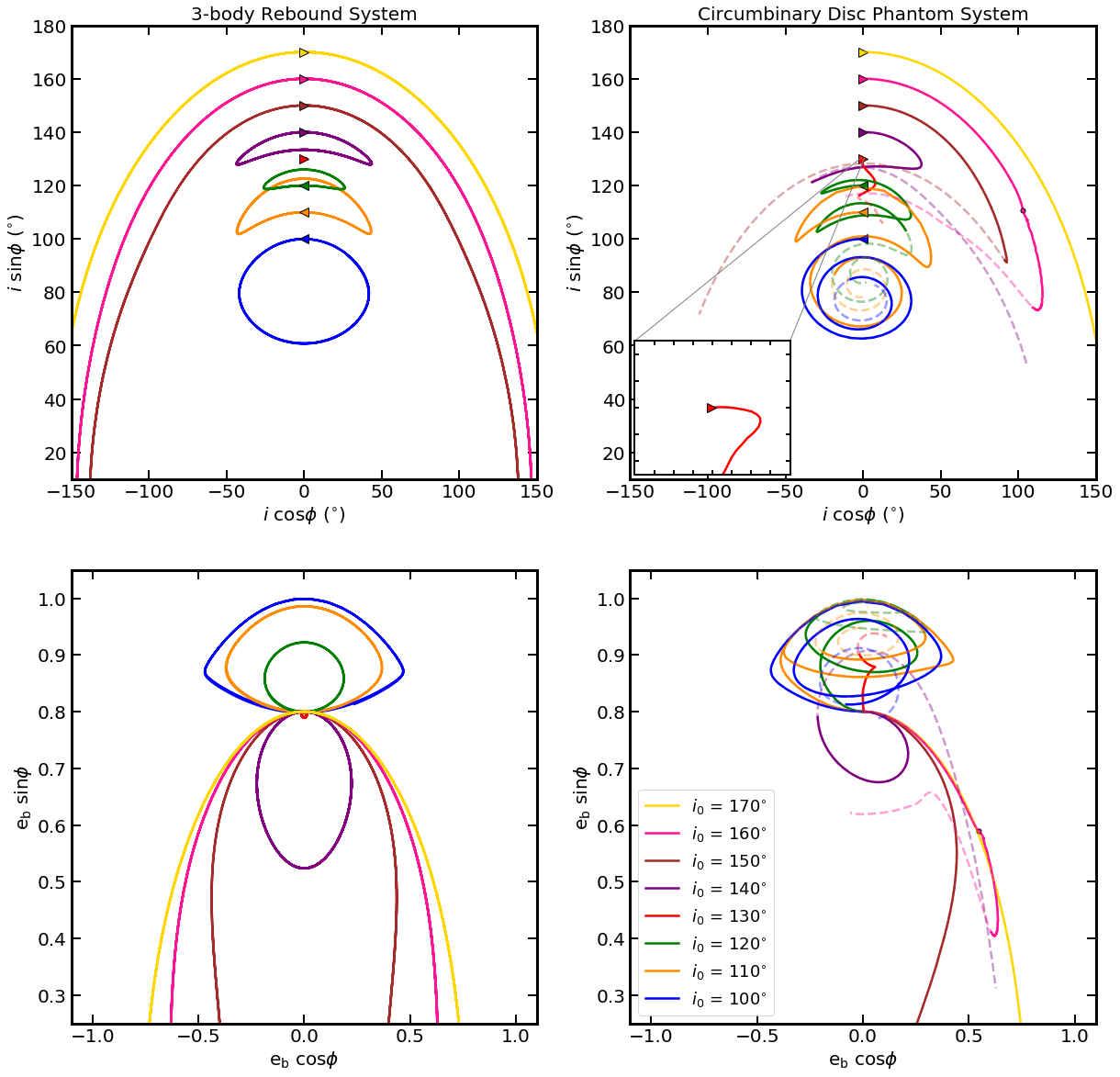}
    \caption{Same as for \ref{fig:md01} except $j=2.5$. The three-body systems have a planet mass $M_{\rm p} = 0.031\,M$ and initial planet distance from system centre of mass $r=20\,a_{\rm b}$. The initial mass of the disc is $M_{\rm d} = 0.05\,M$. }
    \label{fig:md05}
\end{figure*}

We now consider our most massive planet with the highest angular momentum ratio, $j=2.5>j_{\rm cr}=0.91$. The prograde-polar stationary inclination calculated with Equation~(\ref{eq:incstat}) is $i_{\rm s}=70^{\circ}$, and the retrograde-polar stationary inclination is $i_{\rm rs}=129^{\circ}$. The left two panels of Fig.~\ref{fig:md05} show the results of these High-$j$ three-body simulations. The phase plot behaviour is similar to the Mid-$j$ simulations, but with some of the orbital phenomenon shifted downwards by about $10^{\circ}$. For $i_0 \lesssim 100^{\circ}$, the orbit is librating about $i_{\rm s}$. For $i \gtrsim 150^{\circ}$, the orbits are in retrograde circulation. We find the crescent orbit regime to be approximately within $110^{\circ}-140^{\circ}$. Similar to the behaviour of Mid-$j$-140, we find that High-$j$-130 does not deviate significantly from its initial inclination due to its proximity to the retrograde-polar stationary inclination at $i_{\rm rs}=139^{\circ}$. When looking at the eccentricity plots we find the similar behaviours as described in Sec.~\ref{sec:mid_threebody}.

\section{Circumbinary Disc Simulations}
\label{sec:circumbinary}
In this section, we explore hydrodynamical simulations of a circumbinary gas disc with the same initial properties ($i_0$ and $j$) as the three-body simulations in the previous Section in order to understand the long term evolution of a massive circumbinary disc.

\begin{figure*}
	\includegraphics[width=2\columnwidth]{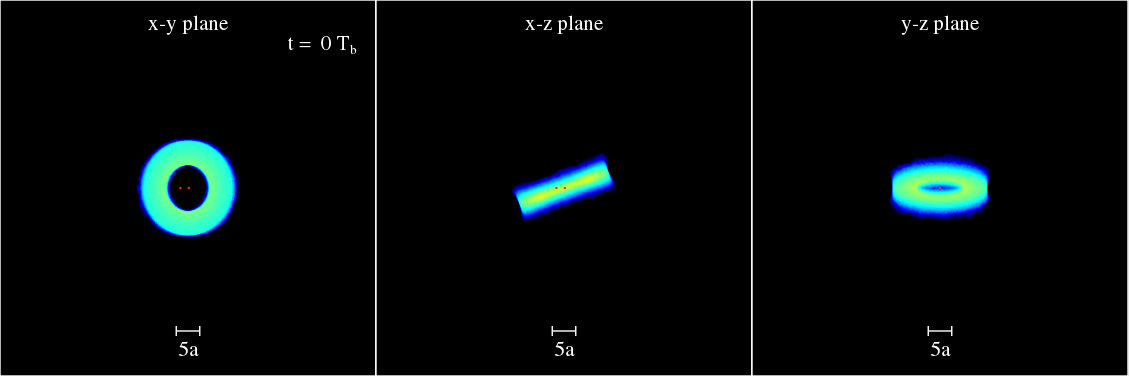}
	\includegraphics[width=2\columnwidth]{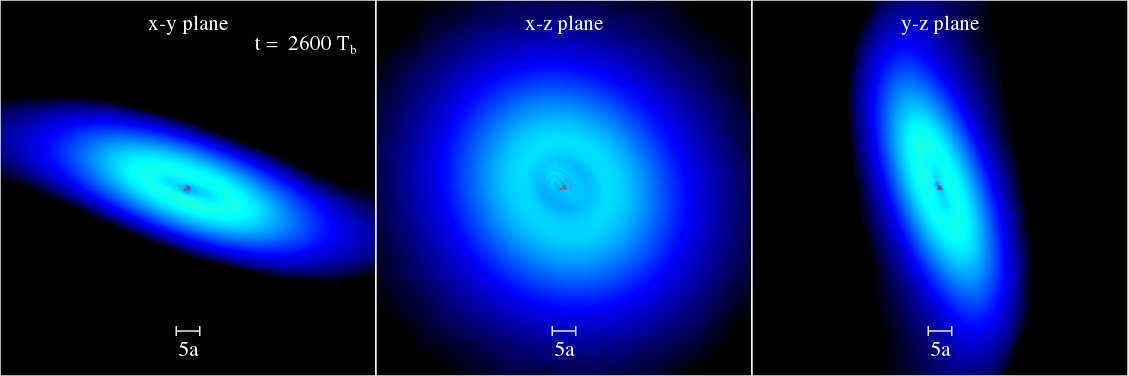}
	\includegraphics[width=2\columnwidth]{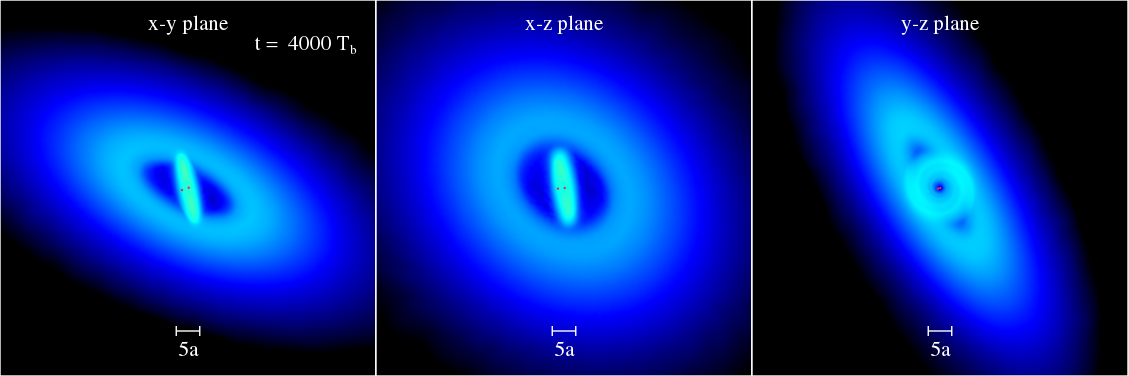}
    \caption{Snapshots from the Low-$j$-160 broken-disc system. The top row shows the initial conditions of the circumbinary disc from three different viewing angles. The middle row shows the time where we  classify the disc as broken and begin analysing the data from the inner and outer disc separately. Note that $t=2600\,T_{\rm b}$ is where the red and blue lines meet in Fig. \ref{fig:broken_disc}. The bottom row shows the simulation at a later time where the two discs are fully separated, and the inner disc has gone nearly polar.}
    \label{fig:splash_all}
\end{figure*}

\subsection{Circumbinary disc simulation set-up}
\label{sec:circumbinary_methods}

We use the smoothed particle hydrodynamics (SPH) code {\sc phantom} \citep{Price2010,Price2018} to simulate circumbinary discs. The binary and disc parameters have the same initial inclination and angular momentum as the three-body simulations in the previous Section. We define the angular momentum ratio as $j = J_{\rm d}/J_{\rm b}$, where $J_{\rm d}$ is the angular momentum of the disc given by
\begin{equation}
    J_{\rm d}=\int_{R_{\rm in}}^{R_{\rm out}} 2\pi r^3 \Sigma(r) \Omega  dr,
\end{equation}
where the Keplerian angular velocity is $\Omega=\sqrt{GM/r^3}$.
All discs start with surface density distributed with a power law $\Sigma \propto R^{-3/2}$ between the initial inner disc radius $R_{\rm in,0} = 5\,a_{\rm b}$ and the initial outer radius $R_{\rm out,0} = 10 \,a_{\rm b}$. The initial inner disc truncation radius is chosen to be initially far from the binary so that the disc expands beyond the initial inner and outer radii. We take the \cite{Shakura1973} $\alpha$ parameter to be $0.01$ in all of our simulations. This disc viscosity is utilized by adapting the SPH artificial viscosity according to \cite{Lodato2010}. The disc is locally isothermal with sound speed $c_{\rm s} \propto R^{-3/4}$ and the disc aspect ratio varies with radius as $H/R \propto R^{-1/4}$. Thus, $\alpha$ and the average smoothing length $\left<h\right>/H$ are constant over the radial extent of the disc \citep[][]{Lodato2007}. We take $H/R = 0.1$ at $R_{\rm in}$. Particles from the disc are removed if they pass inside the accretion radius for each component of the binary at $0.25 \,a_{\rm b}$ \citep{Bateetal1995}. We run our disc simulations with $1\times10^{6}$ particles initially. The disc is initially resolved with averaged smoothing length divided by the disc scale height of $\left<h\right>/H$= 0.15. We have simulations with three disc masses $M_{\rm d} =$ $0.01 \,M$, $0.03 \,M$, and $0.05 \,M$. We run these simulations up until $t = 6000 \,T_{\rm b}$, except for Low-$j$-170, Mid-$j$-140, Mid-$j$-150, High-$j$-130, High-$j$-140, and High-$j$-150 which are run until $t = 10000 \,T_{\rm b}$. However, we indicate the time at which the disc has lost 30\% of its initial mass in both our figures and tables. Beyond this time, the comparison to the equivalent three-body simulations may not be appropriate. The time is shown in the final column in Table~\ref{tab:sim_list}.

One detriment to note in our SPH simulations is that the flow in the central gap region is not well resolved by the code in the intrinsically 3D flows. That flow takes the form of rapid low density gas streams \citep[e.g.][]{Artymowicz1996,Munoz2019,Mosta2019}. This causes some uncertainty in the binary evolution.

In order to analyse the properties of the disc, we divide the disc up into 10000 bins in spherical radius up to a radius of $R_{\rm out} = 100\,a_{\rm b}$. Within each bin we average the orbital properties of the particles, such as the inclination, $i$, and nodal phase angle, $\phi$. Similar to the three-body simulations, we compute the inclination of a ring of the disc at radius $R$ relative to the instantaneous binary angular momentum as
\begin{equation}
    i(R)=\cos^{-1}\left(\boldsymbol{\hat{l}}_{\rm b} \cdot \boldsymbol{\hat{l}}_{\rm d}(R)\right),
	\label{eq:i_bd}
\end{equation}
where $\boldsymbol{\hat{l}}_{\rm d}(R)$ is the unit vector in the direction of the disc angular momentum vector. We then calculate the inclination of the disc as the density weighted average of the inclination 
\begin{equation}
    i=\frac{\int_{R_{\rm in}}^{R_{\rm out}} 2 \pi r \Sigma(r) i(r) dr}{M_{\rm tot}},
\end{equation}
where $\Sigma(R)$ is the surface density at radius $R$, $i(R)$ is the inclination at radius $R$, and $M_{\rm tot}$ is the total mass of the disc at the given time. The longitude of the ascending node phase angle for the disc is
\begin{equation}
    \phi(R)=\frac{\pi}{2}+\tan^{-1}\left( {\frac{\boldsymbol{\hat{l}}_{\rm d}(R) \cdot (\boldsymbol{\hat{l}}_{\rm b} \times \boldsymbol{\hat{e}}_{\rm b})}{\boldsymbol{\hat{l}}_{\rm d}(R) \cdot \boldsymbol{\hat{e}}_{\rm b}}} \right).
	\label{eq:phi_d}
\end{equation}
In a similar fashion to $i$, $\phi$ for the disc is also a density weighted average of the particles in the disc with
\begin{equation}
    \phi=\frac{\int_{R_{\rm in}}^{R_{\rm out}} 2\pi r \Sigma(r) \phi(r) dr}{M_{\rm tot}}.
\end{equation}
In the following Sections, we discuss the results of our circumbinary disc simulations in relation to their corresponding three-body system results.

\subsection{Low-$j$ circumbinary disc}
\label{sec:low_circumbinary}

\begin{figure*}
	\includegraphics[width=1.8\columnwidth]{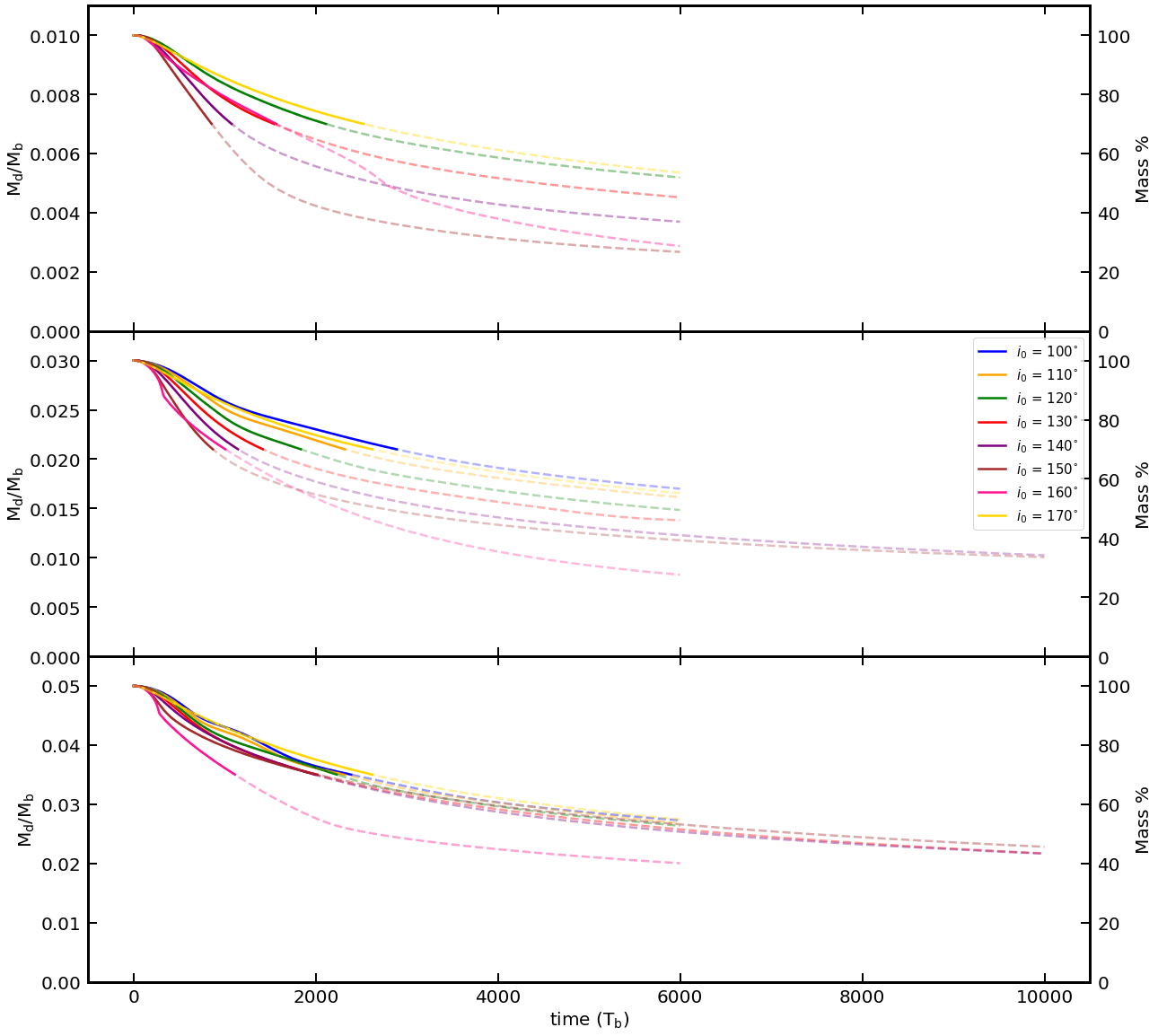}
    \caption{The disc mass evolution of all simulations. The dashed lines represent where each simulation has lost more than $30 \%$ of its initial mass.}
    \label{fig:mass_loss}
\end{figure*}

Here we describe the Low-$j$ hydrodynamical simulations of circumbinary discs, where $j=0.5<j_{\rm cr}=0.91$. The right two plots of Fig.~\ref{fig:md01} show the orbital phase and eccentricity results of these systems. The top right plot shows the $i\cos{\phi} - i\sin{\phi}$ plane and the lower right plot shows the eccentricity phase plot $e_{\rm b}\cos \phi-e_{\rm b}\sin \phi$. When compared to the equivalent three-body system (left), we see general agreement in the phase plots for $i_0\lesssim 160^{\circ}$. While the three-body orbits are exactly periodic, the disc simulations include dissipation which causes the disc to  gradually align toward the prograde-polar stationary inclination  in a spiral-like motion. The upper panel of Fig.~\ref{fig:mass_loss} shows the mass of each disc in time. The lines are dashed after the disc mass has lost 30\% of the initial value. This corresponds to the dashed lines in the phase plots in Fig.~\ref{fig:md01}.

\begin{figure}
	\includegraphics[width=1\columnwidth]{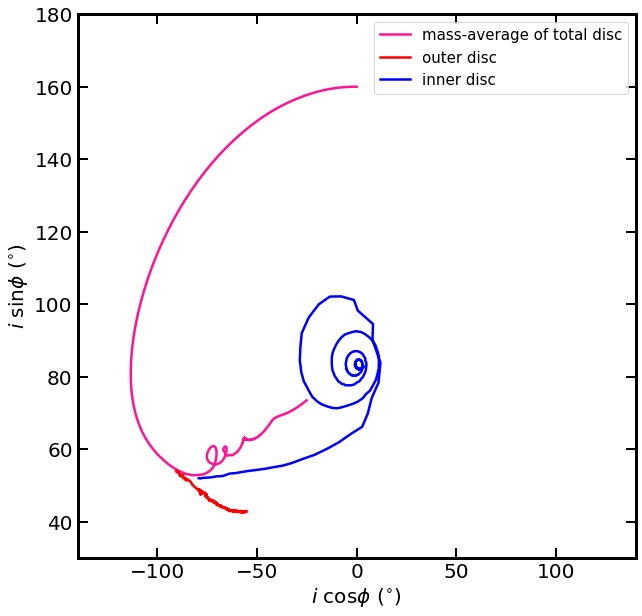}
    \caption{The $i\cos{\phi} - i\sin{\phi}$ phase plot for the Low-$j$-160 broken disc. The blue line represents the inner disc at $R=5\,a_{\rm b}$, while the red line represents the outer disc, set at $R=25\,a_{\rm b}$ after the disc has broken at time of $t=2600\,T_{\rm b}$. The pink line is the same pink line shown in Fig.~\ref{fig:md01}.}
    \label{fig:broken_disc}
\end{figure}

The higher initial inclinations, where $i \gtrsim 160^{\circ}$, show somewhat different behaviour. For Low-$j$-160, the disc initially follows a similar evolution to the particle, but then begins to move quickly towards prograde-polar alignment. This behaviour can be seen in the pink line in Fig.~\ref{fig:md01} which experienced a disc break at a time of $t = 2600 \,T_{\rm b}$. The disc breaks into two disjoint rings at a break radius of around $R = 6 \,a_{\rm b}$. This is visually displayed in Fig.~\ref{fig:splash_all}. Disc breaking occurs when the radial communication in the disc occurs on a longer timescale than the precession timescale \citep{Papaloizouetal1995,Larwoodetal1996,Nixonetal2013}. Protoplanetary discs are typically in the wave-like regime since $H/R \gg \alpha$ \citep{Papaloizou1983} and bending waves, that propagate at about half the sound speed, communicate the warp \citep{Papaloizouetal1995}. Also, Fig.~\ref{fig:broken_disc} shows the inclination phase plot for just this simulation. The pink line here is the same as in the upper right panel of Fig.~\ref{fig:md01}. The blue and red lines show the inner and outer disc respectively plotted. The inner disc is analysed at $R=5 \,a_{\rm b}$, while the outer disc is set at $R=25 \,a_{\rm b}$. The outer disc continues much on the original path, while the inner disc quickly spirals to prograde-polar. The combination of these two paths is why we observe a sharp, but looping path towards prograde-polar alignment. As seen in Fig.~\ref{fig:splash_all}, the radius of the disc break grows over time. Thus, the inclination of the total disc is dominated by the inclination of the inner disc at larger times.

Similar to the three-body paths, we find that the initial direction of the orbital paths (denoted with the coloured triangles) corresponds with the initial direction of the eccentricity plots. The orbital paths with rightward-facing triangles have an initially decreasing $e_{\rm b}$.

\subsection{Mid-$j$ circumbinary disc}
\label{sec:mid_circumbinary}

The right two panels of Fig.~\ref{fig:md03} show the phase plots of the Mid-$j$ circumbinary disc simulations with $j=1.5>j_{\rm cr}=0.91$. When compared to the equivalent Mid-$j$ three-body systems (left), we see general agreement for the highest inclination case of the Mid-$j$-170 system. For $i_0 \lesssim 110^{\circ}$, we see librating precession and the disc spirals towards prograde-polar alignment. The Mid-$j$-110 system in the three-body case shows a crescent orbit but this is not seen for the disc. However, in the approximate range of $120^{\circ}-150{^\circ}$, the disc begins at first on a crescent type orbit. In this range, the disc moves towards libration and prograde-polar alignment. For Mid-$j$-120 and Mid-$j$-130, we see at least one complete phase of a crescent orbit before the disc begins librating. It is important to note that some of these transitions into librating orbits occur after the disc has lost more than $30 \%$ of its initial mass, and thus direct comparison to the three-body system shown may not be applicable. The value for the angular momentum ratio $j$ is decreasing during the simulation because mass is lost from the disc and accreted on to the binary (see the middle panel of Fig.~\ref{fig:mass_loss}). However, for the Mid-$j$-120 simulation, the mass remains high while the orbit transitions to libration.

For Mid-$j$-140 and Mid-$j$-150, we observe the beginning of a descent to prograde-polar alignment which differs greatly from the respective three-body simulations. Note that Mid-$j$-140 begins very close to the retrograde-polar stationary inclination $i_{\rm rs}$ and we see that the initial direction of this orbital path is opposite to that of the respective three-body orbit. We find that most initial inclinations seem to ignore the retrograde-polar stationary inclination $i_{\rm rs}$ in order to evolve towards a librating orbit around prograde-polar alignment $i_{\rm s}$.

Finally, for $i=160^{\circ}$ we again observe the disc breaking at a time of $t=300 \,T_{\rm b}$. We can see the moment of the disc break in the pink inclination and eccentricity paths where there is a small bump and pink dot to indicate the time of the disc break. The approximate radius of this disc break was at $R=2.5\,a_{\rm b}$.

For $i \lesssim 130^{\circ}$, we find that the initial direction of the orbital paths, denoted with the coloured triangles, corresponds with the initial direction of the eccentricity plots. However, for Mid-$j$-140 and Mid-$j$-150 we find the eccentricities initially increasing unlike their respective three-body paths. For Mid-$j$-160, we see an initial decrease followed by an increase in a counter-clockwise rotation unlike the clockwise rotation in the three-body path. For Mid-$j$-170, we see general agreement with both the eccentricity and inclination phase plots.

\subsection{High-$j$ Circumbinary system}
\label{sec:high_circumbinary}

Here we describe simulations of circumbinary discs with $j=2.5>j_{\rm cr}=0.91$. The right two plots of Fig.~\ref{fig:md05} show the results of the High-$j$ circumbinary disc simulations. When compared to the equivalent three-body system, we see general agreement for only the $i_0=100^{\circ}$ and $i_0=170^{\circ}$ systems. For High-$j$-100, we see a librating orbit spiraling towards prograde-polar alignment. For High-$j$-110 and High-$j$-120, we see at least one phase of crescent orbits that descend into librating orbits around prograde-polar alignment. While High-$j$-120 has lost a significant amount of mass before it begins librating, High-$j$-110 has not (see the lower panel of Fig.~\ref{fig:mass_loss}). Specifically for High-$j$-120, we see at least two crescent orbits before descending to prograde-polar libration. For High-$j$-130, we observe the beginning of a descent to prograde-polar alignment which differs greatly from the respective three-body simulation. Note that High-$j$-130 is the closest to the retrograde-polar stationary inclination $i_{\rm rs}$ for $j=2.5$, and we see that the initial direction of this orbital path is opposite to the respective three-body simulation. 

For $i \lesssim 120^{\circ}$, we find that the initial direction of the orbital paths (denoted with the coloured triangles) corresponds with the initial direction of the eccentricity plots (as discussed in Section \ref{sec:high_threebody}). The High-$j$-140 and High-$j$-150 eccentricities initially decrease like their respective three-body paths. For High-$j$-160, we see an initial decrease followed by an increase in a counter-clockwise rotation unlike the clockwise rotation in the three-body path. Similar behaviour can be seen in the pink line in Fig.~\ref{fig:md01} which this system also experienced a disc break at a time of $t=270\,T_{\rm b}$. The disc breaks into two disjoint rings at a break radius of around $R=2.5\,a_{\rm b}$. For High-$j$-170, we see general agreement with both the eccentricity and inclination phase plots.

\section{Conclusions}
\label{sec:conclusion}

We have investigated the evolution of a massive circumbinary disc in order to examine the effect of the disc mass on the probability of polar alignment. We consider a disc that begins with an inclination that is closer to retrograde alignment than prograde alignment to the binary orbit. We have found that at least initially a circumbinary disc displays similar behaviour to a three-body simulation with the same angular momentum ratio of the outer body to the inner binary. Dissipation and radial communication across the disc lead to differences later on. For a narrow range of initial disc inclinations around $i=160^\circ$ the disc communication timescale is longer than the nodal precession timescale and this leads to breaking, where the disc has two disjoint rings that can precess independently. Qualitatively the behaviour is not affected by the disc break.

With three-body simulations we have explored the dynamics of a massive third body orbiting an eccentric binary. There are three different types of nodal precession for massive bodies. First, the third body may be circulating, meaning that it precesses about the binary angular momentum vector (prograde circulation) or the negative of the binary angular momentum vector (retrograde circulation). 
 Second, if the initial inclination is close to the generalised prograde-polar stationary inclination, it may be librating, meaning that it precesses about the generalised prograde-polar state. For inclinations between the retrograde circulation and the libration, the particle may be in a crescent type orbit in the inclination phase plot. These orbits do not have a common centre. There are two stationary inclinations for which there is no nodal precession of the orbit. The prograde-polar stationary inclination lies at the centre of the librating orbits. The retrograde-polar stationary inclination exists  for sufficiently high angular momentum of the third body and lies within the crescent orbits. 
 
 We have found that the dissipation within the circumbinary disc leads to (prograde)-polar alignment for all orbits that begin in the crescent orbit regime. Even if a disc begins very close to the retrograde-polar stationary inclination, it does not stay there. Therefore, discs are not expected to be aligned to the retrograde-polar stationary inclination. Similarly, we do not expect planets to be aligned to the retrograde-polar stationary inclination after the disc has dissipated. Previous models for the probability of polar alignment are based on a massless disc \citep{Aly2015,Zanazzi2018}. However, we have shown that the phase space for which a retrograde disc moves towards polar alignment is larger than would be predicted by just the librating orbits. A massive and close to retrograde circumbinary disc eventually moves towards the prograde-polar stationary inclination. This result has implications for the formation of planets around an eccentric orbit binary since polar discs may form from a wide range of initial disc misalignments.

The existence of crescent type precession orbits and the retrograde-polar stationary inclination relies on a large disc angular momentum (see the critical angular momentum ratio given in equation~\ref{eq:crit_angle}). This corresponds to a disc mass of a few percent or more of the binary mass for a disc that extends to a radius of about 10 times the binary separation that we have considered here.    We now have a number of observational tracers of disc mass that show that disc masses are typically up to a few percent of a solar mass \citep{Anderson2022,Anderson2022b}, some may be up to $0.2\,\rm M_\odot$ \citep{McClure2016}, even larger than those considered in this work. For these disc masses, the range of initial inclinations that evolve to a polar orientation may be large. For example, in Fig.~\ref{fig:md03}, discs with initial inclination in the range $60-150^\circ$ evolve towards polar. Turbulence in giant molecular clouds is the expected way that these disks can form initially misaligned \citep{Bate2018}. If the initial orientation of discs relative to the binary is isotropic, then polar discs may be more common than coplanar discs around eccentric binaries even when then disc is massive. Furthermore, as the disc dissipates and its mass decreases, polar alignment becomes even more likely (see Fig.~\ref{fig:md01} where inclinations $\gtrsim 20^\circ$ evolve towards polar).

\section*{Acknowledgements}
We thank an anonymous referee for useful comments that improved the manuscript. Computer support was provided by UNLV's National Supercomputing Center. We acknowledge support from NASA through grants 80NSSC19K0443 and 80NSSC21K0395. SHL acknowledges visitor support from the Simons Foundation. This research was supported in part by the National Science Foundation under Grant No. NSF PHY-1748958. Simulations in this paper made use of the {\sc rebound} code which can be downloaded freely athttp://github.com/hannorein/rebound. Visualizations of {\sc phantom} simulations were created through Splash from \citep[][]{Price2007}.

\section*{Data availability statement}
The SPH simulations results in this paper can be reproduced using the {\sc phantom} code (Astrophysics Source Code Library identifier {\tt ascl.net/1709.002}). The $n$-body simulation results can be reproduced with the {\sc rebound} code (Astrophysics Source Code Library identifier {\tt ascl.net/1110.016}).  The data underlying this article will be shared on reasonable request to the corresponding author.




\bibliographystyle{mnras}
\bibliography{mnras_template}








\bsp	
\label{lastpage}
\end{document}